\documentclass[a4paper]{jpconf}
\usepackage{amssymb, amsmath, iopams}
\begin{document}

\title{Asymptotic cones and quantum gravity}

\author{N Kalogeropoulos}
\address{Pre-Medical Department, Weill Cornell Medical College in Qatar\\
                   Education City, P.O. Box 24144, Doha, Qatar}

\ead{nik2011@qatar-med.cornell.edu}

\begin{abstract}
       Asymptotic cones are structures that encode  how a metric space appears when seen from far away.
       We discuss their meaning and potential significance for quantum gravity. 
\end{abstract}


\section{Introduction}

 There have been various proposals for the quantisation of gravity alone, each of which has its strengths and weaknesses. Most recently
loop gravity [1], causal dynamical triangulations [2] and causal sets [3] are among the approaches that have attracted attention. 
One of the fundamental beliefs among many quantum gravity approaches is that space-time should be discrete at the Planck scale, in analogy with atoms being the 
fundamental  building blocks of ordinary matter. The different approaches encode such a fundamentally discrete space-time in different ways: loop gravity uses 
holonomies of spin networks and spin foams [1], dynamical triangulations use a priori triangulations of space-time [2] and causal sets start from sets of points 
that are promoted to finite graphs by using the causal structure of space-time [3].  In all of these approaches, recovering classical space-times at appropriate 
semi-classical limits is a strong prerequisite for their viability as physical theories.  For some of these approaches, such as the dynamical triangulations,
obtaining such a semiclassical limit is more straightforward. For some others though, such as loop gravity and causal sets, it is still unclear how to 
generally obtain such a limit, despite considerable efforts in this direction. Classical gravity should arise from its quantised counterpart through  
statistical averaging, at a far larger distance scale than the Planck length. So, one way of approaching the semiclassical limit may be to determine the 
geometric structures that arise if we see the structures of quantum gravity from far away. Asymptotic cones of metric spaces [4-8] are metric spaces themselves that 
express how the underlying metric spaces look when they are seen from far away. As such, they may be of some interest for determining the semi-classical 
limit in some of the afore-mentioned quantum gravity approaches.   

\section{Asymptotic cones: definitions, examples, subtleties}

 Consider $S$ to be a set and denote by \ $\mathcal{P}(S)$ \ to be its power set. A filter \  $\mathcal{F}(S)$ \  is a finitely additive $\{0,1\}$-measure on \ $P(S)$. \ 
The sets that have measure 1 in \ $\mathcal{F}(S)$ \ are assumed to be large. Using Zorn's lemma any filter can be extended to an ultrafilter which can be defined 
by the additional property that for any \ $A\subseteq  S$ \ either \ $A \in \mathcal{F}(S)$ \ or \ $S\backslash A \in \mathcal{F}(S)$. \ The prefix ``ultra" exists because 
such a filter is a maximal element in the ordered set of all filters on \ $S$ \ with respect to inclusion. 
Filters are used extensively in model theory (logic). In the present context, their role is to formally encode the properties that are of significance for our purposes, or 
alternatively, as a book-keeping device. Since we are interested in convergence properties we need the following definition. 
Consider $S$ to be infinite and $a \in S$ and $\mathcal{F}(S) = \{ R \subseteq S: a\in R \}$. Such an ultrafilter is in bijective
correspondence with the elements of $S$ and it is called principal. All other ultra-filters of \ $S$ \ are called non-principal. Non-principal ultra-filters 
generalize statements as the one included in the definition of Cauchy sequences of real numbers: $\forall \ n > n_0 \in \mathbb {N} \ldots$ 
They allow us to focus  in defining limits to the co-finite sets that we deem important for our purposes. At this point a physicist may start becoming a bit uneasy
as there is no constructive description of non-principal ultra-filters, even though we know that they exist by Zorn's lemma.   \\

We are eventually interested in finding continuous structures starting from discrete ones. Hence we want to start by considering sequences of points belonging
to a set $S$ of interest, endowed with additional properties. To be able to compare such sequences, always with a view toward taking limits, 
one needs to choose an ultrafilter \ $\mathcal{F}(S)$. \ Generalising $\mathbb{N}$ which labels elements of usual sequences \ $(x_n), \ n\in\mathbb{N}$, \
one can use an index set \ $I$. \ Let \ $i\in I$. \  Then define the following equivalence relation  
\begin{equation}
   (x_i) \sim (y_i) \ \ \ \mathrm{iff} \ \ \ x_i = y_i, \ \ \ \ \mathcal{F}(S)-\mathrm{almost \ everywhere}
\end{equation}
which generalises the definition of when two sequences \ $(x_n), (y_n),  \ n\in\mathbb{N}$ \ are equal. Indicate such an equivalence class by \ $[x_i]$. \ 
The ultra-product \ $\ast S$ \ with respect to \ $I$ \ and \ $\mathcal{F}(S)$ \ is defined as 
\begin{equation}
    \ast S \ = \ \prod_{\mathcal{F}(S)} \ S \ = \ \prod_{i\in I} S_i \ \slash \sim 
\end{equation}
If all the elements in an ultra-product are the same, then it is called an ultra-power. This is akin to having a copy of \ $S$ \ at every element of  
\ $(x_i)$ \ with identification given by \ $\sim$. \   
As an example of ultra-products we could mention the hyper-reals $\ast \mathbb{R}$ which include the ordinary reals, infinitesimals and 
infinitely large numbers. Ultra-limits are special cases of ultra-products. Endow $S$ with a topology and call  $\mathcal{X}$ the resulting topological space. 
Indicate by \ $\omega$ \ a non-principal ultrafilter on \ $\mathcal{X}$. \ For \ $a\in\mathcal{X}$ \  indicate a neighbourhood of \ $a$ \  by \ $\mathfrak{N}(a)$. \  
An $\omega$ (ultra)- limit is defined as 
\begin{equation}
      \lim_\omega \ (x_n) = a \ \ \ \mathrm{iff} \ \ \ \forall \  \mathfrak{N}(a), \ \ \omega ( \{ n\in\mathbb{N}: x_n \in \mathfrak{N}(a) \}) = 1
\end{equation}
As it can be readily seen this is a straightforward, but non-trivial, generalisation of the Cauchy definition of a limit of a sequence in \ $\mathbb{R}$. \  
One can see, for instance, that \ $\mathbb{R}$ \ is contained in \ $\ast\mathbb{R}$, \ by considering the ultra-limits of the constant sequences $(x_n)$ where \  
$x_n = a\in\mathbb{R}$. \  The $\omega$ (ultra)-limit is unique in any Hausdorff space. 
Asymptotic cones are particular cases of ultra-limits in metric spaces \ $(X, d)$. \ To capture the idea of seeing the space from far away [4] 
we first choose sequences \ $(e_n)\in \ast X$ \ of ``observation centers" or ``base-points" of \ $X$. \  
We also choose a sequence of increasing scaling factors \ $(\alpha_n) \ in \ast\mathbb{R}$ \ such that \ 
$\alpha_n \rightarrow\infty$. \  Let \ $\ast X_e^\alpha$ \ be the set of sequences \ $(x_n)$ \ of points in \ $X$ \ such that \  $d(x_n, y_n) / \alpha_n$ \ 
is bounded and the equivalence relation
\begin{equation} 
   (x_n) \approx (y_n) \ \ \ \mathrm{iff} \ \ \ \lim_{\omega} \frac{d(x_n, y_n)}{\alpha_n} = 0 
\end{equation}
Then define the asymptotic cone of $X$ as \ $Cone_{\omega} X = \ast X_e^\alpha / \approx$. \ Evidently \ $Cone_\omega X$ \ depends on the choice of 
the ultrafilter $\omega$, the scaling factors $(\alpha_n)$ and the observation points $(e_n)$. It is a complete metric space endowed with the metric    
\begin{equation}
       d_\infty ([x_n], [y_n]) \ = \  \lim_\omega \frac{d(x_n, y_n)}{\alpha_n} 
\end{equation}
\\

\noindent To help motivate and clarify the concept of asymptotic cones we present some examples. 
\begin{itemize} 
  \item The asymptotic cone of any bounded space, is obviously, a point.
  \item The Euclidean space \ $\mathbb{R}^n$ \ has as asymptotic cone \ $\mathbb{R}^n$ \ itself: 
             indeed \ $\mathbb{R}^n$ \ lacks a characteristic  length  scale.
 \item   The Abelian group $\mathbb{Z}^n$ endowed with its word metric has asymptotic cone \ $\mathbb{R}^n$ \ endowed with the 
             ``Manhattan/taxicab" metric \ $d(x,y) = |x_1-y_1| + \ldots |x_n - y_n|$. This is obvious: as someone walks further and further away from \ 
            $\mathbb{Z}^n$ \ the distance between its points decreases and from an infinite distance it becomes indistinguishable from \ $\mathbb{R}^n$.
             The limit metric is similarly obvious [4].  
 \item A generalisation:  a finitely generated group is virtually Abelian if its asymptotic cones are isometric to the Eucidean space \ $\mathbb{R}^n$ [4]. 
 \item Further generalisation: A finitely generated group is virtually nilpotent if and only if its asymptotic cones are locally compact [4],[5].
\end{itemize}
 The above examples include spaces that are either flat or almost flat or behaving
  as if they are almost flat, in a sense. These results are independent of the ultra-filter considered and do not actually need the machinery of ultra-limits 
 to be used. The usual pointed Gromov-Hausdorff convergence is sufficient [9]. Ultra-filters are used because without them one could not even 
 formulate the question of what is the asymptotic cone of the hyperbolic space $\mathbb{H}^n$. The problem is that as one goes to infinity, 
 the boundary of  $\mathbb{H}^n$ is ``too big" as seen by the linear isoperimetric inequality for $\mathbb{H}^n$ [9]. The pointed Gromov-Hausdorff 
 convergence relies on uniform nets and cannot handle such an exponential increase of volume and area. 
 \begin{itemize}
  \item Using ultra-filters one can see that $Cone_\omega \mathbb{H}^n$ is a topological tree, each point of which has degree the continuum \ $\aleph_0$ [9].  
  \item For Gromov hyperbolic spaces the same is true: their asymptotic cones are trees each point of which has degree $\aleph_0$ [7].
  \item The asymptotic cone of a non-compact symmetric space of rank at least two, is a (non-discrete) Euclidean building [10].
\end{itemize}

\noindent There are several important points and subtleties related to asymptotic cones. To wit
\begin{itemize}
   \item If one chooses different ``observation centre" sequences that are a finite distance apart, then the corresponding asymptotic cones turn out to be isometric. 
   \item The choice of ultra-filters and dilation factors is far more crucial though: different choices of ultra-filters may lead to non-homeomorphic asymptotic cones [11].
   \item The question of how many non-isometric cones does a ``simple space" such as a finitely generated group have, has a ``disturbing" partial answer [12]:
             If the continuum hypothesis is not true then any uniform lattice (namely a finite co-volume discrete subgroup) of \ $SL_n(\mathbb{R})$ \ has \ 
             $2^{2^{\aleph_0}}$ non-isometric asymptotic cones. If the continuum hypothesis is true, then any such lattice has a unique asymptotic cone.               
\end{itemize}
The last statement is somewhat unnerving because, according to a particular viewpoint, a statement about nature cannot possibly be related to 
fundamental (or even worse: undecidable) propositions in set theory or logic, foundation problems in quantum gravity non-withstanding.
Even though ultra-filters are at a confluence of logic and geometry that allows quite general constructions, encountering the validity of propositions such 
as the continuum hypothesis in a physical setting has implications that need to be addressed and justified.\\

It is generally perceived that determining the asymptotic cones of metric spaces is a difficult task as they tend to be ``large" and ``indescribable" [13]. So we may have 
do with less. Determining the topology of such spaces, for instance, may be a useful start. In such cases one may be able to at least develop topological field 
theories on them in an attempt to make a connection with Physics. Or someone could be interested in determine large-scale topological defects arising from 
a more conventional field theory. For the case of quantum gravity such ``defects" may play a fundamental role, geons [14], which may also help address the
issue of topology change of space-time [15]. But even if we try to determine the topology of asymptotic cones we run into serious difficulties. One example mentioned 
above implies that the vast majority of asymptotic cones are not expected to be even locally compact. Another case is a theorem stating that there exists a finitely 
generated group $G$ with continuously many non $\pi_1$-equivalent asymptotic cones, irrespective of the validity of the continuum hypothesis [13]. To do better and 
attempt to tame the wilderness of the possibilities and ``pathologies" that seem to arise with asymptotic cones, we may have to impose stronger conditions on the 
underlying space. A theorem in this direction states that groups, behaving like a Euclidean space, namely having a quadratic isoperimetric function have simply 
connected asymptotic cones [16]. Several results like this exist in the literature, but their possible use for quantum gravity is, currently, unclear.\\

It may appear from the above, that our insistence in extensively using discrete groups as examples of possibly interesting metric spaces is unjustified and 
artificial, from a physical viewpoint. We have two reasons for doing so. 
The first is economical: groups are probably the simplest algebraic structures that have the concept of successive measurements, encoded 
via the composition/multiplication of their elements, built-in. Alternatives certainly exist; most notably categories have attracted some attention recently, 
but the point still remains for the interest in the case of groups. 
Second, if one wishes to work in the Hamiltonian framework in $3+1$ gravity, thus excluding possible topology changes of space-time, a critical 
element is an understanding of the isometry classes of space-like hypersurfaces and the implementation of the diffeomorphism constraint on them. 
Such hypersurfaces are 3-dimensional (Riemannian) manifolds.  As such, and after a performing a prime and/or a Jaco-Shalen-Johansson decomposition [17],     
the metric structure of each building block has a strong group theoretical flavour: each is an orbifold whose cover possesses one of the eight geometries  
in Thurston's geometrization program [18], which was more recently confirmed by Perel'man [19]. It is not  clear on whether such results exist for  
the subset of 4-dimensional topological manifolds admitting metrics of Lorentzian signature or even degenerate metrics.


\section{Conclusions and future directions} 

The above are some preliminary thoughts about the potential use of asymptotic cones in quantum gravity. We have pointed out some of the subtleties and some 
examples of these constructions. These structures have a strong hyperbolic geometric flavour. It is unclear how they can be adapted to work in spaces of variable 
sign curvature, although the more recent work on tree-graded spaces and relatively hyperbolic groups may start providing an answer [13]. Another important question 
is how fast is the convergence of a space to its asymptotic cone(s): what is the ``time-scale" the thermodynamic limit is reached [20]? Naturally the bigger issue
is whether these structures can be of any use in quantum gravity. The above hand-waving arguments can be interpreted as indicating that they may be promising. \\

It should be noted that all the mathematical ideas, examples etc. contained in the present work are well-known, even classical.
Our goal was to bring such constructions to the attention of the gravitational physics community. The reference list is very incomplete due to space 
limitations.  

\ack{We wish to thank the organizers of \ IC-MSQUARE 2013 \ for inviting us to present this work.}


\end{document}